\documentclass[aps,pra,showpacs,showkeys,showkeys,amsmath,amssymb]{revtex4}
\usepackage{graphicx}
\usepackage{dcolumn}
\usepackage{bm}

\def\vep{\varepsilon}
\def\ben{\begin{equation}}
\def\een{\end{equation}}
\def\bea{\begin{eqnarray}}
\def\eea{\end{eqnarray}}

\begin{document}
\title{Bose-Einstein Condensation of Confined
and Non-interacting Bose Particles by the Integral
Representation of Bose Functions}
\author{Sang-Hoon \surname{Kim}
\footnote{shkim@mmu.ac.kr} }
\affiliation{Division of Liberal Arts, Mokpo National Maritime
University, Mokpo 530-729, Korea}
\date{\today}
\begin{abstract}
With the integral representation of Bose functions,
 the Bose-Einstein condensation of non-interacting bosons
 in a three-dimensional harmonic trap was studied.
The relation between the particle number and
its phase transition temperature was clarified.
Some next-order terms in the thermodynamic expansions were obtained.
We plotted the chemical potential, the mean energy,
and the specific heat and found
most of these properties obtained by using the integral representation
were almost identical with those of
the series representation of Bose functions.
\end{abstract}
\pacs{05.30.Jp, 03.75.Fi, 32.80.Pj}
\keywords{Bose-Einstein condensation,
Condensate fraction, Specific heat}
\maketitle

After the historical paper by Bose and Einstein in the 1920s, the
observation of a Bose-Einstein condensation (BEC) was successfully
demonstrated in magneto-optical traps (MOT) of dilute alkali gases
about a decade ago \cite{ande,davi}. Although the Bose atoms
in the MOT  are confined and interacting \cite{kim,kim1,kim2}, a
confined and non-interacting Bose system was studied because it is
analytic and gives some clue for the lab system when the number of
particles is not very large. However, the previous studies have been
based on series representations of the Bose functions
\cite{gros,mull,haug1,haug2}.

In this paper, we study a very fundamental part of the BEC of
confined and non-interacting Bose particles by using an integral
representation of the Bose functions \cite{robi}.
We obtain the relation
between  the number  of particles and the shift of $T_c$, the
temperature of the phase transition, exactly. The chemical potential
and the specific heat around $T_c$ will be displayed to find the phase
transition. In principle, we will follow the method introduced by
Haugerud {\it et al.} during the statistical analysis
\cite{haug1,haug2}.

Consider a 3D system of $N$ non-interacting bosons confined by a
harmonic potential of angular frequency ${\bf \omega}$.
The discrete
one-body energy eigenvalues of the harmonic oscillator
potential are given by
$\vep_{n} = \hbar(n_x \omega_x+n_y\omega_y+n_z\omega_z)+\vep_0, $ where
$n_x,n_y,n_z=0,1,2,...$ and $n=\{n_x,n_y,n_z\}$.
The zero-point
energy is given by $\vep_0 = (1/2)\hbar(\omega_x +\omega_y +\omega_z)$.
The distribution function of bosons in the grand canonical ensemble
at temperature $T$ is given by
\ben
N=\sum_{n_x,n_y,n_z}
\frac{1}{e^{\beta(\vep_n -\mu)}-1},
\label{10}
\een
where
$\beta=1/k_B T$, $k_B$ is the Boltzmann constant, and
$\mu$ is the chemical potential.
For simplicity, we here  ignore the anisotropy of the trap and
consider a typical trap potential of
$\omega_x=\omega_y=\omega_z=\omega=10^3 sec^{-1}$.
	Since each energy level has a degeneracy
$d(n)=(n+1)(n+2)/2$, the total number $N$ in Eq. (\ref{10}) is written as
\bea
N &=& \frac{1}{e^{\beta(\vep_0 -\mu)}-1}
+ \sum_{n=1}^{\infty} \frac{d(n)}{e^{\beta(\vep_n -\mu)} -1}
\nonumber \\
&=&\frac{1}{z^{-1} e^{\beta\vep_0}-1} + \sum_{m=0}^{\infty}
\frac{\frac{1}{2}m^2+\frac{5}{2}m+3 }
{e^{-\beta\mu+\beta\hbar\omega+\beta\hbar\omega m} -1},
\nonumber \\
&  \equiv & N_0 + N_e,
\label{20}
\eea
where $z=e^{\beta \mu}$ is the fugacity.

There are various ways to convert the summation into integrals, and
here we use the Euler-Maclaurin's summation-integral formula
\cite{mull,haug1,haug2,abra}. Then, the $N$ in Eq. (\ref{20}) turns
into the following integral forms:
 \bea N &=& N_0 +
\frac{1}{(\beta\hbar\omega)^3}F_3(\alpha) +
\frac{5}{2(\beta\hbar\omega)^2}F_2(\alpha)
\nonumber \\
& & + \frac{3}{\beta\hbar\omega}F_1(\alpha) + \frac{31}{24}\frac{1}{
e^{\alpha} -1 } +\frac{\beta \hbar \omega }{4} \frac{
e^{\alpha}}{(e^{\alpha} -1 )^2},
\label{40}
\eea
where
$\alpha=-\beta\mu + \beta\hbar\omega$,
$e^{-\alpha}=e^{\beta(\mu-\hbar\omega)} = z_e $ is the effective
fugacity \cite{haug1,haug2}, and
 $F_s(\alpha)$ is the integral
representaion of Bose functions defined by \cite{robi}
\ben
F_s(\alpha) \equiv \frac{1}{\Gamma(s)} \int_0^\infty
\frac{x^{s-1}}{e^{\alpha+x} -1} dx,
\label{50}
\een
with $ s > 0$.
$F_s$ is related to the series representation of Bose functions, $
g_s(z) = \sum_{n=1}^\infty z^{n}/n^s $, by way of the effective
fugacity as $F_s(\alpha)=g_s(z_e)$.

The Riemann-zeta function is a special case of the integral as
$ F_s(0)=g_s(1)=\zeta(s)$, and there are several more useful properties \cite{abra}:
\ben
\frac{d F_s(\alpha)}{d \alpha} = -F_{s-1}(\alpha),
\label{57}
\een
where $s > 1$.
Note that $d g_s(z)/dz = g_{s-1}(z)/z$.
When $\alpha$ is small,
$F_s(\alpha)$ can be expanded by using a power series
of the $\alpha$ as \cite{robi}
\bea
F_s(\alpha) &=&
\Gamma(1-s) \alpha^{s-1} + \sum_{n=0}^\infty \zeta(s-n) \alpha^n
\nonumber \\
&\simeq& \Gamma(1-s) \alpha^{s-1} + \zeta(s) + \dots ~
\label{58}
\eea
For the first two values of the integer $s$, $F_s(\alpha)$ is given by
\bea
F_2(\alpha) &=& \zeta(2) -\alpha (1-\ln \alpha) + \dots ~,
\\
F_1(\alpha) &=& -\ln \alpha + \dots ~
\label{60}
\eea
	Since $ \beta\hbar\omega \ll 1$, $N_e$ in Eq. (\ref{40}) is expanded
around $ -\beta \mu $ with the help of Eq. (\ref{57}) as
\bea
N_e &=&
\frac{1}{ ( \beta\hbar\omega )^3 } \left[ F_3( -\beta \mu ) + F'_3(
-\beta \mu ) \beta\hbar\omega + \frac{1}{2!}F''_3(-\beta \mu )
(\beta\hbar\omega)^2  \right]
\nonumber \\
& & + \frac{5}{2 ( \beta\hbar\omega )^2 }
\left[ F_2( -\beta \mu ) + F'_2( -\beta \mu ) \beta\hbar\omega  \right]
+\frac{3}{  \beta\hbar\omega } F_1( -\beta \mu )
\nonumber \\
& & + \frac{31}{24}\frac{1}{ e^{-\beta \mu }-1 }
+\frac{1}{4} \frac{\beta \hbar \omega e^{-\beta \mu }}{(e^{-\beta \mu } -1 )^2}
+ \dots
\nonumber \\
&=& \frac{1}{(\beta\hbar\omega)^3}F_3(-\beta\mu ) +
\frac{3}{2(\beta\hbar\omega)^2}F_2(-\beta\mu) + {\cal O} \left(
\frac{1}{\beta\hbar\omega} \right).
\label{70}
\eea
Substituting Eq.
(\ref{70}) into Eq. (\ref{20}), we find that the number of particles in
the condensate  varies with the temperature as
\cite{gros,mull,haug1,haug2}
\ben
N_0  = N - \left( \frac{k_B
T}{\hbar\omega}\right)^3 F_3(-\beta\mu) - \frac{3}{2}\left(
\frac{k_B T}{\hbar\omega}\right)^2 F_2(-\beta\mu).
\label{80}
\een
In the transition region, $\mu \rightarrow 0$,  then, $F_s$ becomes
the Riemann-zeta function $\zeta(s)$.

Defining the transition temperature $T_c$ where $N_0 = 0$
in Eq. (\ref{80}), we find that
\bea
T_c &=& \frac{\hbar\omega}{k_B}\left(  \frac{N}{\zeta(3)} \right)^{1/3}
\left[ 1 - \frac{3\zeta(2)}{2 N}
\left( \frac{k_B T_c}{\hbar\omega}\right)^2 \right]^{1/3}
\nonumber \\
&=& T_0 \left[
1-\frac{A}{2} ( 1 - A + A^2 - A^3 +  \dots) \right]
\nonumber \\
&=& T_0 \left[ 1 - \frac{A}{2(1 + A)}\right], \label{90} \eea where
\ben T_0 = \frac{\hbar \omega}{k_B} \left( \frac{N}{\zeta(3)}
\right)^{1/3}. \label{92} \een This is the microcanonical critical
temperature for the infinite system. $A$ in Eq. (\ref{90}) is the
one-dimensional inverse particle number defined by \ben A(N) \equiv
\frac{\zeta(2)}{\zeta(3)^{2/3}N^{1/3}} = \frac{1.455}{N^{1/3}}.
\label{94} \een
	Therefore, the ratio of the shift of  $T_c$ is given by \cite{gros}
\ben S(N) = \frac{T_0 - T_c}{T_0} = \frac{A}{2(1+A)}. \label{900}
\een If $N \le 198$, then the $S(N)$ is more than $10 \%$.
Similarly, $N=2000$ corresponds to about $5\%$ and $N=20000$ to about
$2.5\%$.
We plotted the $S(N)$ as a function of $N$ in Fig. 1. Since
$S(N) \rightarrow 0$ for large $N$, $T_0$ will be a useful standard
temperature for discussion. Therefore, Eq. (\ref{80}) is written
with the definition of $T_0$ as \cite{haug1,haug2}
\ben
\frac{N_0}{N} = 1 - \left( \frac{T}{T_0} \right)^3
\frac{F_3(-\beta\mu)}{\zeta(3)} - \frac{3 }{2}\left(
\frac{T}{T_0}\right)^2 \frac{F_2(-\beta\mu)}{\zeta(3)^{2/3}
N^{1/3}}.
\label{93}
\een
The condensate fraction is plotted in Fig. 2 for increasing $N$
The sudden flattening of the curves takes place around  $T_c$.

The negative shift \cite{gros}  in Eq. (\ref{90}) is expanded
up to the order of $N^{-2/3}$ as
\bea
\frac{T_c -T_0}{T_0}
&=& -\frac{A}{2(1+A)}
\nonumber \\
&=& -\frac{0.7275}{ N^{1/3}} + \frac{1.059}{N^{2/3}} + \dots ~
\label{100}
\eea
The chemical potential for $N=200$, $N=2000$, and
$N=20000$ is plotted in Fig. 3.
It approaches 0 as $T \rightarrow T_c$.
Note that there is a cross at $T=T_0$.

The specific heat is obtained from the first derivative of its mean
energy. The mean energy of the system is the internal energy of $N$
non-interacting and harmonically confined bosons:
\ben U =
\sum_{n=0}^\infty \frac{d(n) \vep_n}{e^{\beta(\vep_n - \mu)}-1}.
\label{120}
\een
The zero-point mean energy is small in the
condensate state.
If we apply the Euler-Maclaurin formula Eq. (\ref{120}),
we can write $U(T)$ in the similar way with Eq.
(\ref{70}) as
\bea \frac{U}{\hbar\omega} &=& \frac{N \vep_0
}{\hbar\omega} +\sum_{m=0}^\infty \frac{\frac{1}{2}m^3 + 3 m^2 +
\frac{11}{2}m + 3 } {e^{\alpha + \beta\hbar\omega m - \mu}-1}
\\
&=& \frac{U_0}{\hbar\omega}
+ \frac{3}{ (\beta\hbar\omega)^4 } F_4(\alpha)
+ \frac{6}{ (\beta\hbar\omega)^3 } F_3(\alpha)
+ \frac{11}{2(\beta\hbar\omega)^2} F_2(\alpha)
\nonumber \\
& &+ \frac{3}{\beta\hbar\omega} F_1(\alpha)
+ \frac{25}{24} \frac{1}{e^{\alpha}-1}
+ \frac{\beta\hbar\omega}{4} \frac{e^{\alpha}}{(e^{\alpha}-1)^2} + \dots ~
\label{130}
\eea

We may expand $U$ around $\alpha=-\beta\mu$. Then, we have
\ben
\frac{U}{\hbar\omega} = \frac{U_0}{\hbar\omega} + \frac{3}{
(\beta\hbar\omega)^4 } F_4(-\beta\mu) + \frac{3}{ (\beta\hbar\omega)^3 }
F_3(-\beta\mu) + \frac{1}{(\beta\hbar\omega)^2} F_2(-\beta\mu) +\dots~
\label{135}
\een
Note that  the $F_1$ term does not exist here.
Replacing $k_B/\hbar\omega$ with $T_0$ in Eq. (\ref{92}), we obtain
the energy per particle  as a function of $T/T_0$:
\ben
\frac{U-U_0}{N\hbar\omega} = 3 \left(\frac{T}{T_0} \right)^4
\frac{F_4(-\beta\mu)}{\zeta(3)^{4/3} N^{-1/3}} + 3 \left(
\frac{T}{T_0} \right)^3 \frac{F_3(-\beta\mu)}{\zeta(3)} + \left(
\frac{T}{T_0} \right)^2 \frac{F_2(-\beta\mu)}{\zeta(3)^{2/3}
N^{1/3}}.
\label{137}
\een
We plotted $(U-U_0)/N\hbar\omega$ of Eq.
(\ref{137}) in Fig. 4. The phase transition is shown at $T_c$.

If we take the partial derivatives with respect to temperature in
Eq. (\ref{135}), the specific heat around $T_c$ is obtained for two
ranges:
\\
(i)
When $T \le T_c$,
\ben
\frac{C_V^-}{N k_B} = 12  \left( \frac{T}{T_0} \right)^3
\frac{F_4(-\beta\mu)}{\zeta(3)} + 9  \left( \frac{T}{T_0} \right)^2
\frac{F_3(-\beta\mu)}{\zeta(3)^{2/3} N^{1/3}}.
\label{160}
\een
(ii)
When $T > T_c$,
\bea
\frac{C_V^+}{N k_B} &=& 12 \left( \frac{T}{T_0}
\right)^3 \frac{F_4(-\beta\mu)}{\zeta(3)} + 9 \left( \frac{T}{T_0}
\right)^2 \frac{F_3(-\beta\mu)}{\zeta(3)^{2/3}N^{1/3}}
\nonumber \\
& & - 3 T_0 \left[ \left( \frac{T}{T_0} \right)^4 \frac{F_3(-\beta\mu)}{\zeta(3)}
+\left(  \frac{T}{T_0} \right)^3
\frac{F_2(-\beta\mu)}{\zeta(3)^{2/3} N^{1/3}} \right]
\frac{\partial \alpha}{\partial T}.
\label{210}
\eea
The $ \partial \alpha /\partial T$ is obtained from
Eq. (\ref{92}) as
\ben
\frac{\partial \alpha}{\partial T} =
\frac{3}{T} \frac { F_3(-\beta\mu ) + \frac{T_0}{T} \left(
\frac{\zeta(3)}{N} \right)^{1/3} F_2(-\beta\mu )} {F_2(-\beta\mu ) +
\frac{3T_0}{2T} \left( \frac{\zeta(3)}{N} \right)^{1/3}
F_1(-\beta\mu)}.
\label{215}
\een
When $\mu$ approaches zero,
$\partial\alpha/\partial T$ in Eq. (\ref{215}) becomes zero, too,
because $F_1$ diverges.
Therefore, $C_V^-$ in  Eq. (\ref{160}) and
$C_V^+$ in Eq. (\ref{210}) match at $T_c$. The basic procedure is
the same as shown in Ref. \cite{gros}.

$C_V/N k_B$ is plotted in Fig. 5. The specific heat
exhibits a sharp jump like the $\lambda$ transition.
Since the divergence of $F_1$ is extremely slow,
it is very difficult to find
the continuity at $T_c$ in the numerical calculation, but the sudden
drop of $C_V$ in the vicinity of $T_c$ is clear.
As previous research pointed out \cite{gros,haug1,haug2}, the $\lambda$-like
phase transition in the specific heat is seen for any small number of
non-interacting and confined bosons.

The discontinuity of the specific heat at $T_0$ was obtained as a
function of $N$. If we expand this equation up to  terms of
$1/N^{1/3}$, we obtain the following form:
 \ben
\frac{C_V^+}{N k_B} =
12 \left( \frac{T}{T_0}\right)^3  \frac{ F_4(-\beta\mu)}{\zeta(3)} -
9 \left( \frac{T}{T_0}\right)^3 \frac{ F_3(-\beta\mu)^2}{\zeta(3)
F_2(-\beta\mu)}.
\label{220}
\een
Except  the $T_c$, the point of sudden drop of the slops,
the divergence of $F_1(-\beta\mu)$ is
extremely slow.
Therefore, at $T > T_c$, the $N^{1/3}$ dominantes
$F_1$ and the approximation is obtained as
\bea
 \frac{C_V^- -
C_V^+}{N k_B} & =& 9  \frac{\zeta(3) }{\zeta(2)} + 9 \left(
\frac{\zeta(3)}{N} \right)^{1/3}
\nonumber
\\
&=& 6.577 +
\frac{9.569}{N^{1/3}}.
\label{230}
\eea
 If $N$ is order of $10^3$, the second term  is about 10$\%$ of the first term.

We rewrote the BEC of  confined and non-interacting Bose particles
in 3D by using the integral representation of Bose functions. We
obtained the exact relation between the number of particles and the
shift of $T_c$ and showed how the thermodynamic limit of the
condensate fraction is achieved around $T_c$ with increasing $N$.
The chemical potential around  $T_c$ was plotted for increasing $N$.
	It has a common point at $T_0$, regardless of the number of particles.
We plotted the mean energy and the specific heat around
$T_c$. Although we used the integral representation of the Bose
functions, we see that the results are almost identical with
previous results obtained by using a series representation of Bose functions.

The author sends many thanks to S. Gnanapragasam and M. P. Das for
useful discussions.

{}
\pagebreak
\begin{figure}
\includegraphics{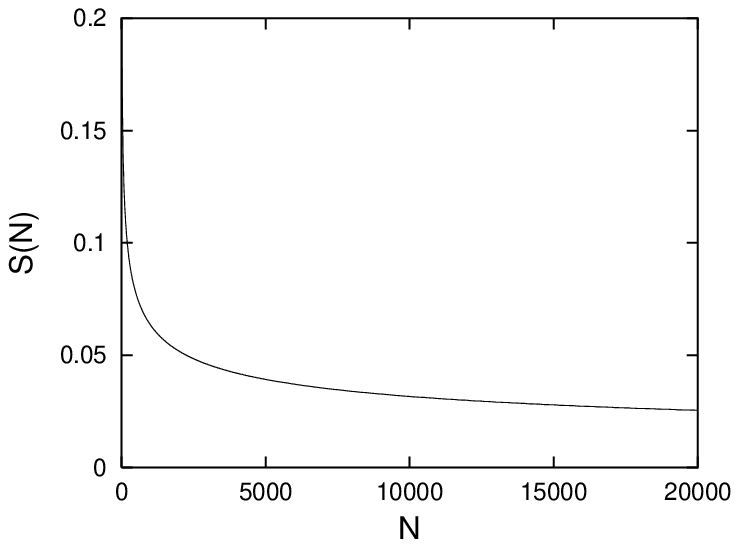}\\
\caption{ Ratio of the shift of the phase transition temperature
as a function of $N$.}
\end{figure}

\begin{figure}
\includegraphics{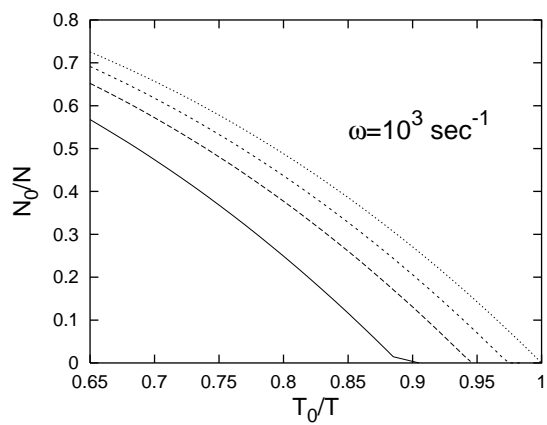}\\
\caption{ Condensate fraction around $T_c$.
The solid line is for $N=200$,
the long-dashed line is for $N=2000$,
the short-dashed line is for $N=20000$, and
the dotted line is for the $N \rightarrow \infty$ limit. }
\end{figure}

\begin{figure}
\includegraphics{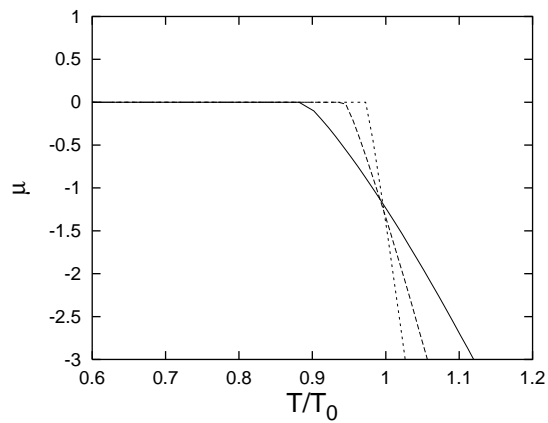}\\
\caption{ Chemical potential around $T_c$.
The solid line is for
$N=200$, the long-dashed line is for $N=2000$, and
the short-dashed line is for $N=20000$. }
\end{figure}

\begin{figure}
\includegraphics{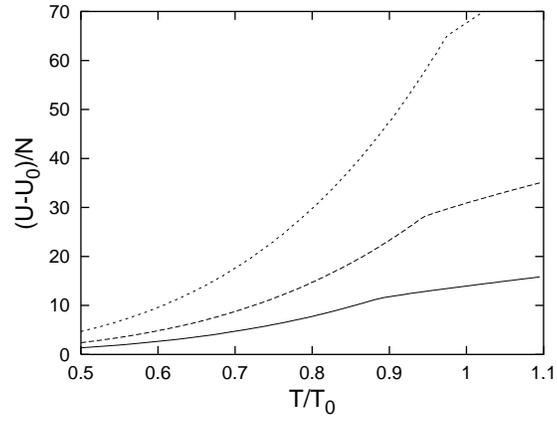}\\
\caption{Mean energy around $T_c$. The unit is $\hbar\omega$.
The solid line is for $N=200$,
the long-dashed line is for $N=2000$, and
the short-dashed line is for $N=20000$. }
\end{figure}

\begin{figure}
\includegraphics{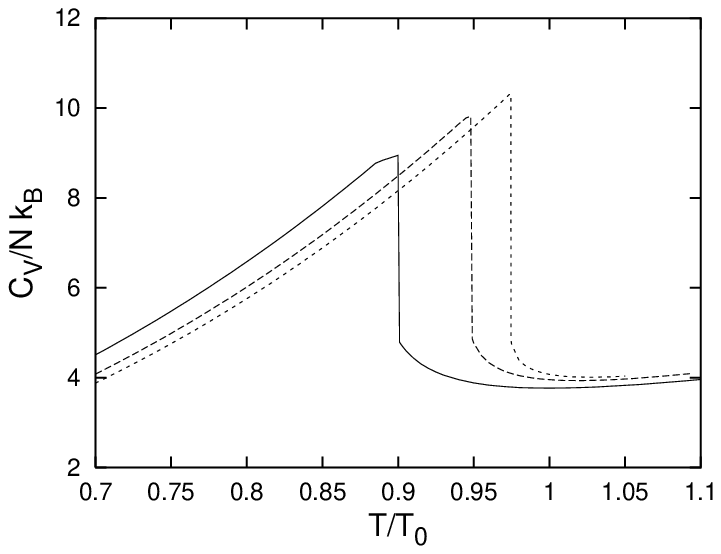}\\
\caption{ Specific heat around $T_c$. From left to right,
$N=200$, $N=2000$, and $N=20000$. }
\end{figure}

\begin{thebibliography}{}
\bibitem{ande} M. H. Anderson, J. R. Ensher, M. R. Matthews, C. E. Weimann,
and E. A. Cornell, Science {\bf 269}, 198 (1995).
\bibitem{davi} K. B. Davis, H.-O. Mewes, M. R. Andrews, N. J. van Druten,
D. S. Durfree, D. M. Kurn, and W. Ketterle, \prl {\bf 75}, 3969 (1995).
\bibitem{kim} S.-H. Kim, S. D. Oh, and W. Jhe, J. Korean Phys. Soc.
{\bf 37}, 665 (2000).
\bibitem{kim1} D. Kim, G. Jin, and J.-H. Yoon  J. Korean Phys. Soc.
{\bf 46}, 1336 (2005).
\bibitem{kim2} S.-H. Kim and C. S. Kim, J. Korean Phys. Soc.
{\bf 48}, 158 (2006).
\bibitem{gros} S. Grossmann and M. Holthaus,
 Phys. Lett. A {\bf 208}, 188 (1995).
\bibitem{mull} W. J. Mullin,  J. Low Temp. Phys. {\bf 106}, 615 (1997).
\bibitem{haug1} H. Haugerud, T. Haugset, and F. Ravndal,
 Phys. Lett. A {\bf 225}, 18 (1997).
\bibitem{haug2} T. Haugset, H. Haugerud, and J. O. Anderson, \pra
{\bf 55}, 2922 (1997).
\bibitem{robi} J. E. Robinson,  Phys. Rev. {\bf 83}, 678 (1951).
\bibitem{abra} M. Abramowitz and I. A. Stegun,
{\it Handbook of Mathematical Functions} (Dover, New York, 1972).
\end{thebibliography}
\end{document}